\pdfoutput=1 \hoffset=0.1in \voffset=-0.3in
\documentclass[11pt]{article}
\usepackage{fancyhdr}
\usepackage{amsfonts}
\usepackage{amsmath,amssymb,amsthm}

\usepackage{mathrsfs}
\usepackage{caption}
\usepackage{subcaption}

\usepackage[sort&compress,numbers,merge]{natbib}
\usepackage{graphicx}
\newcommand{\FMDG}[1]{\includegraphics{#1}}

\usepackage{multirow}
\usepackage{color}
\usepackage[colorlinks]{hyperref}
 \hypersetup{
 pdfauthor={Yasaman Farzan, Silvia Pascoli and Michael A. Schmidt},
 pdftitle={Recipes and Ingredients for Neutrino Mass at Loop Level}
 }

\makeatletter
\renewcommand{\fnum@table}{\textbf{\tablename~\thetable}}
\renewcommand{\fnum@figure}{\textbf{\figurename~\thefigure}}
\makeatother

\newcommand{\eV}{\ensuremath{\,\mathrm{eV}}}

\newcommand{\TeV}{\ensuremath{\,\mathrm{TeV}}}

\newcommand{\braket}[1]{\ensuremath{\left<#1\right>}}
\newcommand{\ev}[1]{\ensuremath{\left<#1\right>}}
\newcommand{\hc}{\ensuremath{\text{h.c.}}}

\newcommand{\be}{\begin{equation}}
\newcommand{\ee}{\end{equation}}
\newcommand{\ba}{\begin{eqnarray}}
\newcommand{\ea}{\end{eqnarray}}

\newcommand{\Eqref}[1]{Eq.~(\ref{#1})}

\newcommand{\Figref}[1]{Fig.~\ref{#1}}
\newcommand{\Tabref}[1]{Tab.~\ref{#1}}
\newcommand{\Secref}[1]{sec.~\ref{#1}}

\renewcommand{\thefootnote}{\fnsymbol{footnote}}

\begin{document}
\allowdisplaybreaks[1]

%%%%%%%%%%%%%%%%%%
%%  TITLE PAGE  %%
%%%%%%%%%%%%%%%%%%

\begin{titlepage}
\ \vspace*{-15mm}
\begin{flushright}
 IPM/P-2012/027\\
IPPP/12/64\\
DCPT/12/128
\end{flushright}
\vspace*{5mm}

\begin{center}
{\Large \sffamily\bfseries
 Recipes and Ingredients for Neutrino Mass at Loop Level }\\[10mm]

{\large
Yasaman Farzan\footnote{\texttt{yasaman@theory.ipm.ac.ir}}$^{(a)}$,
Silvia Pascoli\footnote{\texttt{silvia.pascoli@durham.ac.uk}}$^{(b)}$, and
Michael A.~Schmidt\footnote{\texttt{michael.schmidt@unimelb.edu.au}}$^{(c)}$}
\\[5mm]
{\small\textit{$^{(a)}$ School of Physics, Institute for research
in fundamental sciences (IPM), P.O. Box 19395-5531, Tehran, Iran
}}\\
{\small\textit{$^{(b)}$
Institute for Particle Physics Phenomenology (IPPP), University of Durham,
Durham DH1 3LE, UK
}}\\
{\small\textit{$^{(c)}$
ARC Centre of Excellence for Particle Physics at the Terascale,
School of Physics, The University of Melbourne, Victoria 3010, Australia
}}

\end{center}
\vspace*{1.0cm}
\date{\today}

\begin{abstract}
The large hierarchy between the neutrino mass scale and that of
the other fermions seems to be unnatural from a theoretical point
of view. Various strategies have been devised in order to generate
naturally small values of neutrino masses. One of these techniques
is neutrino mass generation at the loop level which requires a
mechanism, {\it e.g.,} a symmetry,  to forbid the lower order
contributions. Here, we study in detail the conditions on this
type of symmetries. We put special emphasis on the discrete $Z_n$
symmetries as a simple example but our results can be also
extended to more general groups. We find that regardless of the
details of the symmetry, in certain cases the existence of a lower
order contribution to neutrino masses can be determined by the
topology of the diagrams with a given number of loops. We discuss
the lepton flavor violating  rare decays as well as $(g-2)_\mu$
 in this class of models, which generically appear
 at the one loop level. Typically the imposed symmetry has important implications for dark matter, with the possibility of stabilizing one or even multiple dark matter candidates.
\end{abstract}

\end{titlepage}

\newpage
\setcounter{footnote}{0}
\renewcommand{\thefootnote}{\arabic{footnote}}

\section{Introduction}

Explaining  nonzero but tiny neutrino masses is one of the most compelling open
questions in modern physics. Various beyond Standard Model (SM)
theories have been developed to address this question.
The most famous mechanism for explaining neutrino masses is the
standard (type I) seesaw
mechanism~\cite{Minkowski:1977sc,*Yanagida:1980,*Glashow:1979vf,*Gell-Mann:1980vs,*Mohapatra:1980ia}.
In the standard and simplest realization of this mechanism, the
smallness of neutrino masses is connected to the very large mass
scale of new SM singlets (right-handed (RH) neutrinos). These new
particles, being too heavy, cannot be produced at the LHC or any
other man-made or natural environment, (maybe) except for the
early universe, making a direct test of these models impossible.
With the start of the LHC data release, it is more exciting to
move towards models whose new particles are within the reach at
the LHC.  The smallness of the neutrino masses is not related
anymore uniquely to the very heavy mass scale of the RH neutrinos,
but requires additional suppressions, {\it e.g.,} small Yukawa couplings,
quasi-conserved lepton symmetries. A very interesting possibility
is to forbid neutrino masses at tree level and have them generated
at loop-level. The first proposals of this type of radiative
neutrino mass models are the Zee model at one
loop~\cite{Zee:1980ai} and the Zee-Babu model at two
loop~\cite{Zee:1985rj,*Zee:1985id,*Babu:1988ki}.

 All seesaw
models lead to the  effective dimension 5 operator, $(HL) (HL)$.
However in general, neutrino masses are not necessarily
explained by this dimension 5 operator. Various other $\Delta L=2$
operators can also give rise to neutrino mass. These operators have
been classified in~\cite{Babu:2001ex,*de_Gouvea:2007xp,*Angel:2012ug}. A
particular class of these operators are $(H^\dagger H)^m (HL)(HL)$ with more than one pair of Higgs fields attached to the
corresponding diagram \cite{Babu:2009aq,*Bonnet:2009ej,*Kanemura:2010bq,*Liao:2010ku}.
Recently, there has been a complete classification of one loop
diagrams leading to the effective dimension 5 operator in
\cite{Bonnet:2012kz} following earlier work~\cite{Babu:1989fg,*Ma:1998dn}. Obviously, the loop generated neutrino
masses receive further corrections from renormalization group
running, which have been studied for Ma's scotogenic model
in~\cite{Bouchand:2012dx}.

 Let us suppose
that, thanks to a specific structure of the model, up to the
n$^\mathrm{th}$ loop-level, there is no contribution to the
neutrino mass matrix. Of course, increasing the loop order will
further suppress the neutrino mass. At the three-loop order, with
$M_{NEW}\sim 100$ GeV and couplings of order of 0.1, by
dimensional analysis the neutrino mass will be in the range
${\mathcal{O}}\left( (g^2/16 \pi^2)^3
M_{NEW}\right)-{\mathcal{O}}\left( (g^2\log( \Lambda/M_{NEW})^2/16
\pi^2)^3 M_{NEW}\right)\sim (0.01 -1)~\eV$ with a cutoff scale
$\Lambda\sim (1-10)\TeV$. Such a high value of coupling and low
mass scale is very interesting from a phenomenological point of
view as it can lead to observable effects in colliders and
indirect searches of new physics. Furthermore, the couplings
leading to the neutrino mass can also in principle induce Lepton
Flavor Violating (LFV) rare decays and a contribution to the
anomalous magnetic moment of the muon.
The construction of these radiative seesaw models often requires the introduction of an
additional symmetry, $G_\nu$, forbidding the tree-level
contribution as well as contributions from lower loop orders. An
interesting consequence of these symmetries is that they stabilize
some of the new degrees of freedom and these models can provide a suitable dark matter (DM)
candidate~\cite{Krauss:2002px,Cheung:2004xm,*Asaka:2005an,*Kubo:2006yx,*Chun:2006ss,*Hambye:2006zn,*Kubo:2006rm,*Babu:2007sm,*Gu:2008yj,*Sahu:2008aw,*Aoki:2008av,*Aoki:2009vf,*Hambye:2009pw,*Ma:2009gu,*Adulpravitchai:2009re,*MarchRussell:2009aq,*Okada:2010wd,*Li:2010rb,*Meloni:2010sk,*Hirsch:2010ru,*Adulpravitchai:2010wd,*Kanemura:2011vm,*Chang:2011kv,*Lindner:2011it,*Kanemura:2011mw,*Ahn:2012cga,*Chao:2012sz,Ma:2006km,Boehm:2006mi,Farzan:2009ji,Farzan:2010mr}.
Besides a $Z_2$ parity, there are several studies involving larger symmetry groups~\cite{Ma:2007gq,Batell:2010bp,Adulpravitchai:2011ei,Belanger:2012vp}.

In this paper, we will restrict ourselves to models which lead to
the effective dimension 5 operator. Hence, the left-handed lepton
doublets $L$ are the only fermions coupling to the new particles,
{\it i.e.} leptons act as a portal to the hidden sector. Employing
a $G_\nu$ symmetry, we  restrict  the couplings to the form
$\mathcal{L}_Y=L S_i F_j$, where $S_i$ ($F_j$) are new scalars
(fermions) and forbid couplings of the form $L H F_j$ as well as
$LL S_i$. In this context, we will be general and will not
restrict ourselves to the content of a specific model. Our aim is
to outline general restrictions and no-go-theorems as a guide to
build radiative neutrino mass models. We will consider radiative
neutrino mass models up to three loop order.   Beyond three loop
order, the induced neutrino mass is becoming too small to explain
the atmospheric neutrino mass scale. We demand that all the SM particles are invariant
under the new symmetry, $G_\nu$, but some or all of the new
particles transform under $G_\nu$. In particular, we assume all
new particles that couple directly to $L$ as well as all the new
neutral fermions carry a $G_\nu$ charge. Moreover, we assume that
none of the new scalars receives a vacuum
 expectation value: that is, $G_\nu$ remains unbroken.
 As a result,  the $G_\nu$ symmetry forbids a Dirac mass term
 for the SM neutrinos both at the tree level and at all orders of
 perturbation theory.
 The neutrino mass
 term should be  therefore of Majorana type which in turn requires  lepton number violation.

For concreteness, we first consider Abelian $Z_n$
symmetries ({\it i.e.,} we take $G_\nu=Z_n$).  We classify the emerging topologies up to three loop order and
discuss the conditions on the $Z_n$ symmetry which forbid all lower
loop orders. We then show that
some of the results we find also
hold valid for a $U(1)$ symmetry and more general symmetries $G_\nu$ in \Secref{sec:generalisation}.

The paper is organized as follows. In \Secref{sec:general}, we
outline the general setting of the models discussed in the present
paper and some general implications for neutrino masses. In
\Secref{sec:neutrino}, we discuss the loop contributions to the
neutrino mass matrix and show how the $Z_n$ symmetry can forbid
lower order contribution to the neutrino mass. In
\Secref{sec:generalisation}, we discuss how the $Z_n$ symmetry can
be generalized to other groups.   In \Secref{sec:lfv}, we discuss
the restrictions from LFV rare decays and anomalous magnetic
moment of the muon. In \Secref{sec:DM}, we briefly
discuss the implications of the $G_\nu$ symmetry for dark matter. In \Secref{sec:con}, we summarize our conclusions
and briefly comment on implications for LHC signatures.

%%%%%%%%%%%%%%%%%%%%%%%%%%%%
\section{General setting of the models\label{sec:general}}
%%%%%%%%%%%%%%%%%%%%%%%%%%%%%%

We extend the Standard Model by introducing $N_S$ scalars, $S_i$, $i=1, \dotsc, N_S$, and $N_F$ left-handed Weyl fermions, $F_j$, $j=1, \dotsc, N_F$. We assume that the leptons constitute a portal to the hidden sector via a Yukawa coupling of form
%%%
\be {\mathcal{L}}_Y=\sum_i^{N_S}\sum_j^{N_F}g_{ij\alpha} S_i F_j
L_\alpha,  \ \ \ \mbox{where} \ \ \ L_\alpha=\left(\begin{matrix}
\nu_\alpha \cr l_\alpha^- \end{matrix}\right)\label{L-Y} ~. \ee
Throughout the paper, we will adopt a two component notation and
write all fields as left-handed Weyl spinors, { \it i.e.} in the
$(\frac12,\,0)$ representation. The product of two Weyl spinors
$\xi, \,\eta$ is therefore defined by $\xi\eta\equiv\xi^Tc\eta$,
where $c$ denotes the charge conjugation matrix ({\it i.e.,}
$c_{11}=c_{22}=0$; $c_{12}=-c_{21}=1$). We will use an index-free
notation, unless a special discussion of the Lorentz structure is
required. Notice that only the combinations of form $F_j L_\alpha$
in \Eqref{L-Y} are allowed. A combination of form $F_j^\dagger
L_\alpha$ is forbidden by the Lorentz symmetry.  We focus on the
neutrino mass generation via coupling  \Eqref{L-Y}  and   do not
consider other
 radiative neutrino mass generation mechanisms, {\it e.g.,}
  via a coupling of
the new particles to RH charged leptons (See e.g. \cite{Krauss:2002px})
 or via two $W$
boson exchange~\cite{Babu:1988ig}.

Of course, ${\mathcal{L}_Y}$ should be a $SU(2)\times U(1)$
invariant combination. As a result, the sum of hypercharges of
$S_i$ and $F_j$ is opposite to the hypercharge of $L_\alpha$, {\it
i.e.,} $ Y_{S_i}+Y_{F_j}=-Y_{L_\alpha}=1$. In case that the
hypercharge of the chiral fermions that we are adding is non-zero,
anomaly cancellation might require addition of extra chiral
fermions. There are various ways to make the combination invariant
under $SU(2)$. For example, if $F_j$ is a triplet and $S_i$ is a
doublet,  the combination $\epsilon_{ab} S_i^a (F_j)^{bc}
L_\alpha^c$ is a $SU(2)$ invariant  ($a$, $b$ and $c$ are $SU(2)$
indices). Our discussion of the  implications of the $Z_n$
symmetry  for loop contributions to the neutrino mass matrix is
independent of the behavior of the fields under $SU(2)$ so we
shall not specify the  behavior of the fields under the
electroweak symmetry.

Taking $G_\nu=Z_n$, the fields in \Eqref{L-Y}
transform as follows
\begin{eqnarray} S_i &\to& e^{i\frac{2\pi}{n}\alpha_{S_i}} S_i ~,\cr
 F_j&\to& e^{i\frac{2\pi}{n}\alpha_{F_j}} F_j ~,\cr
 L_\alpha &\to& L_\alpha  ~, \end{eqnarray}
 such that
 \be
\alpha_{S_i}+\alpha_{F_j}\equiv 0\mod n\quad\Leftrightarrow\quad
 \frac{\alpha_{S_i}+\alpha_{F_j}}{n}\in
 {\mathbb{Z}}\quad\Leftrightarrow\quad
 \alpha_{S_i}+\alpha_{F_j}\in n\mathbb{Z}
 .\label{principal}
 \ee
 If we promote $Z_n$ to $U(1)$, the condition in
\Eqref{principal} should be replaced by
$\alpha_{S_i}+\alpha_{F_j}=0$.

Obviously, a Majorana mass term for the fermion $F$ is
forbidden, unless the $Z_n$ charge of $F$ fulfils
$2\alpha_{F} \equiv 0 \mod n$. A Weyl fermion $F$ with
$2\alpha_{F} \neq 0\mod n$ therefore needs another Weyl fermion,
$F^\prime$ with $\alpha_F=-\alpha_{F^\prime}$ to form a Dirac mass
term. In case that $F$ and  the conjugate of $F^\prime$ are in the
same representation of $SU(2)\times U(1)$, the mass term will be
simply of form $F F^\prime$. Anomaly cancellation in this framework
will be automatic. If $F$ and $F^\prime$ are in different
representations of $SU(2)\times U(1)$, anomaly cancellation might
require additional chiral fields and a mass term can emerge only
after electroweak symmetry breaking. For example, if $F$ is a
doublet and $F^\prime$ is a singlet, the mass term can originate
from a term of form $F F^\prime H$.

%%%%%%%%%%%%%%%%%%%%%%%%%%%%%%%%%%%%%%%
%%%%%%%%%%%%%%%%%%%%%%%%%%%%%%%%%%%%%%%%%%%%
\section{Loop contributions to neutrino masses \label{sec:neutrino}}
%%%%%%%%%%%%%%%%%%%%%%%%%%%%%%%%%%%%%%%
%%%%%%%%%%%%%%%%%%%%%%%%%%%%%%%%%%%%%%%%%
As discussed earlier, we focus on models within which a Dirac
neutrino mass is forbidden by a $Z_n$ symmetry  and Majorana
neutrino masses are produced only at loop level. In subsection
\ref{general-remarks}, we make general remarks on the loop
contributions to the neutrino mass. In subsection
\ref{sec:OneLoop}, we focus specifically on the one-loop
contribution. In subsection \ref{theorem}, we discuss the
conditions for constructing a lower loop contribution to the
neutrino mass using the
 propagators and vertices in
 a general multi-loop diagram, with a specific discussion about
 the two-loop and three-loop cases in the
subsequent subsections. We will analyze the different possible
topologies without specifying the SM model charges or even the
number of new fields.

\subsection{ General remarks on the loop level neutrino
mass\label{general-remarks}}

We consider diagrams contributing to neutrino masses and we will indicate the scalar
propagators by dashed lines and the fermion propagators
 by solid ones.
In general, the scalar propagator can involve
more than one scalar: $\braket{S_1 S_2^\dagger}$. Notice that a
propagators of form $\braket{S_1 S_2}$ can be rewritten in form of
$\braket{S_1 S_3^\dagger}$ by
redefining $S_3\equiv S_2^\dagger$.
Without loss of generality, we will work in a basis
with diagonal kinetic terms;
as a result, a propagator of form
$\braket{F_1 F^\dagger_2}$ for $F_1\ne F_2$ does not exist.
The propagators of form $\braket{F F^\dagger}$  preserve any
 $Z_n$ symmetry. In general,
the fermionic propagator can be either chirality flipping ({\it i.e.,} of form $
\langle F_1 F_2^T c\rangle$ or $ \langle cF_1^* F_2^\dagger
\rangle$) or chirality conserving ({\it i.e.,} of form  $\langle F F^\dagger \rangle$).
Propagators of type $\braket{F_{1} F_{2}^Tc}$ and
$\braket{cF_{1}^* F_{2}^\dagger}$ can result from Dirac or
Majorana mass terms.

In order to generate a Majorana mass term $\nu\nu$ for neutrinos, lepton number has to be
broken by two units.
This can be achieved in various ways. An extensively studied option is
to have a Majorana mass term for the new fermions and
a lepton number violating mass term of form $m^2 S^2/2$ for the new scalars.  However, the options in general are wider. For example
in case of the two-loop diagram in \Figref{fig:2loop.a}, if we
assign lepton number equal to $-1$ to $S_1$ and $S_2$ (or to $F_1$
and $F_2$), lepton number will  be broken by two units by the
$S_1S_2S_3$ vertex or the $F_1 F_2 S_3^\dagger$ vertex.

A Majorana neutrino mass term $\nu\nu$  can arise only after
electroweak symmetry breaking. As we already briefly mentioned,
the Weinberg operator is the lowest dimension operator that can
induce Majorana mass for neutrinos. Higher dimension operators can
also result in neutrino mass. See
\cite{Babu:2001ex,*de_Gouvea:2007xp} for a classification of all
$\Delta L=2$ operators leading to neutrino masses. However,
additional SM fields, which are not charged under $G_\nu$ do not
affect our discussion of the $G_\nu$ symmetry besides generating
the neutrino mass at a higher loop order. A particular class of
these operators are $(H^\dagger H)^m (HL)(HL)$ with
more than one pair of
 Higgs fields attached to the diagram, which have been studied
  in \cite{Bonnet:2009ej,*Kanemura:2010bq,*Liao:2010ku}.
In the following, we will concentrate on the simplest origin and
only discuss the Weinberg operator $(HL)(HL)$, {\it i.e.,}~a pair
of Higgs $H$ being attached to the loop diagram giving mass to neutrinos. Let us
discuss each option separately.
\begin{itemize}
\item

The Higgs can be attached to a vertex of type $S_1 S_2 S_3$  via a
renormalizable coupling $H S_1 S_2 S_3$, provided that this
combination forms a singlet. However, it cannot be attached to a
fermionic vertex because the corresponding term in the Lagrangian
will be non-renormalizable.

 \item
 Let us now discuss the case in which the Higgs field is attached
  to the propagators. Propagators of form $\langle F F^\dagger\rangle$ and $\langle S
S^\dagger \rangle$ cannot break $SU(2)\times U(1)$ but propagators
of type $\langle S S\rangle$ or $\langle F F^Tc\rangle$ in
principle can do so. Let us take a general propagator of form
$\langle \phi \psi^\dagger \rangle$ where $\phi$ and $\psi$  are either both  scalars or  both left-handed fermions.
 If $\langle \phi \psi^\dagger\rangle$ breaks
hypercharge by one (two) units, electric charge conservation
implies $T_3(\phi)-T_3(\psi)=1~(2)$. This means that $\phi$ and
$\psi$ cannot be both  singlets. A propagator involving only one
Dirac field in an $SU(2)$ doublet representation can break hypercharge by at most one unit. However, if
we allow more than one field to be involved in the propagators,
more possibilities open up.  The line shown in
 \Figref{fig:DiracMassInsertion} is an example.
\begin{figure}
\begin{center}
\resizebox{0.3\linewidth}{!}{\FMDG{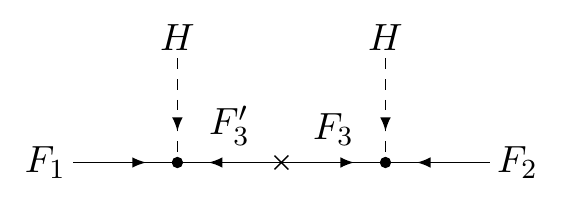}}
\end{center}
\caption{Fermion propagator breaking the hyper-charge by two units
\label{fig:DiracMassInsertion}}
\end{figure}
To have such a line, the required terms are $HF_{3}F_{2}$, $H
{F}_{3}^\prime F_{1}$ and $m_{F_3}{F}_{3}F_{3}^\prime$. From now
on, for brevity we shall not emphasize on the requirement of
mass insertion for a line such as the one denoted by $F_3$ and
$F'_3$ in \Figref{fig:DiracMassInsertion}.
\end{itemize}
Consider a loop that contains $n_I$ internal lines plus $n_{SV}$
vertices that involve three scalars. For such a diagram, there are
$$n_I\left[(n_I+1)/2+n_{SV}\right]+n_{SV}(n_{SV}-1)/2$$ ways to
attach the pair of external Higgs fields to the diagram. For
example, in case of \Figref{fig:3loopPlanar.b} $n_I=8$ and
$n_{SV}=3$ which means there are 63  ways of attaching the
external pair of Higgs fields. To avoid cluttering the figures with this
plethora of possibilities, we do not show the Higgses attached to
the diagrams.

%%%%%%%%%%%%%%%
\subsection{One loop\label{sec:OneLoop}}
\begin{figure}
\begin{center}
\begin{subfigure}{0.3\textwidth}
\FMDG{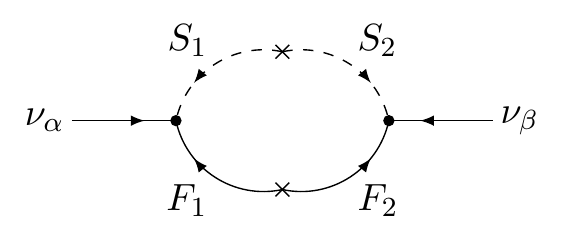}
\caption{One loop diagram with mass term $SS$ on scalar line\label{fig:OneLoopMaj}}
\end{subfigure}
\hspace{2cm}
\begin{subfigure}{0.3\textwidth}
\FMDG{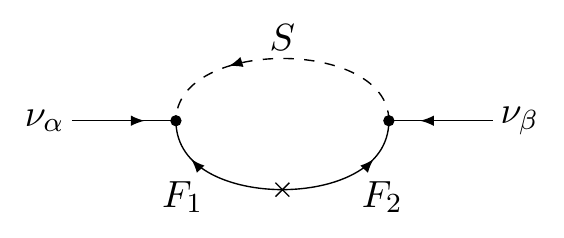}
\caption{One loop diagram without mass term $SS$ on scalar line.\label{fig:OneLoopNoMaj}}
\end{subfigure}
\end{center}
\caption{Effective neutrino mass generation at one loop.\label{fig:OneLoop}}
\end{figure}

%%%%%%%%%%%%%%%%%%%%%%%%%
At one loop, there are two possible diagrams, which are not
accompanied by a tree-level contribution. They are shown in
\Figref{fig:OneLoop}. Had we included neutral fermions
invariant under $G_\nu$ or allowed new scalars to develop a VEV,
we could have more types of one-loop diagrams
\cite{Bonnet:2012kz}.

The propagators in the one-loop diagram \Figref{fig:OneLoopMaj}
are of form $\langle S_1 S_2\rangle$ and $\langle F_{1}
F_{2}^Tc\rangle$ where $S_2$ and $F_2$ may or may not be the same
as $S_1$ and $F_1$, respectively. Let us denote the $Z_n$ charge
of an arbitrary field $\phi$ by $\alpha_\phi$. In order for the
propagators to be $Z_n$ invariant, the charges of $S_1$ and $S_2$
as well as the ones of $F_1$ and $F_2$ have to add up to $0\mod
n$. The existence of the vertices leads to similar conditions. The
following relations need to be satisfied
\begin{equation}
\alpha_{S_1}+\alpha_{S_2}, \,\alpha_{F_1}+\alpha_{F_2},\,\alpha_{S_1}+\alpha_{F_1},\, \alpha_{S_2}+\alpha_{F_2}\in n\mathbb{Z} ~,
\end{equation}
which lead to
\begin{equation}
\alpha_{S_1}\equiv -\alpha_{S_2}\equiv \alpha_{F_2}\equiv -\alpha_{F_1}\mod n ~.
\end{equation}
 In the specific case that the pair of $(S_1, F_1)$ is identified with
 $(S_2,F_2)$, we find $2\alpha_{S_1}\equiv2\alpha_{F_1}
\equiv\alpha_{S_1}+\alpha_{F_1}\equiv 0 \ {\rm mod} \   n$ and if there is no other field $\phi$ with
non-trivial $Z_n$ parity, any choice for $n$ will be
equivalent to $n=2$. However, in general when $S_1 \ne S_2$ and
$F_1 \ne F_2$, $n$ might be different from 2.

Let us discuss the special case $S\equiv S_1 =S_2^\dagger$, which
corresponds to the diagram shown in \Figref{fig:OneLoopNoMaj}.
Apparently, in this case the scalar line, $\langle S S^\dagger
\rangle$ cannot break $SU(2)\times U(1)$ which means both Higgs
fields have to be attached to the fermion line.  As discussed
earlier, if the fermion line is needed to break hypercharge by two
units, it has to involve at least one fermion in addition to $F_1$
and $F_2$ (see \Figref{fig:DiracMassInsertion}). Notice that in
this case no lepton number violating mass insertion of type
$m^2S^2/2$ is required. Instead, the simultaneous presence of  the
$S F_1L$ and $S^\dagger F_2L$ vertices and the $F_1 F_2$ mass term
breaks lepton number. The Lorentz structure of the Majorana mass
term, $\nu\nu$, cannot be created by a fermion propagator of type
$\langle F F^\dagger\rangle$ so a mass insertion of the fermionic
propagator is required.
\begin{figure}
\begin{center}
\begin{subfigure}{0.3\textwidth}
\FMDG{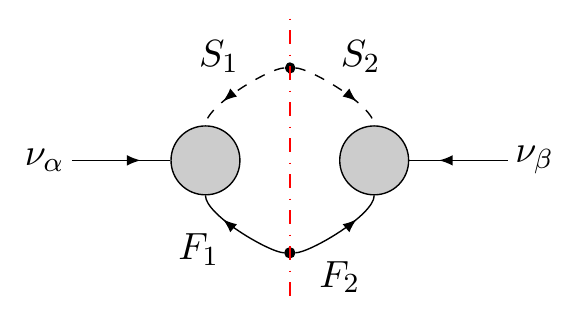}
 \caption{ \label{fig:GeneralLoopDiagram}}
\end{subfigure}
\hspace{1cm}
\begin{subfigure}{0.3\textwidth}
\FMDG{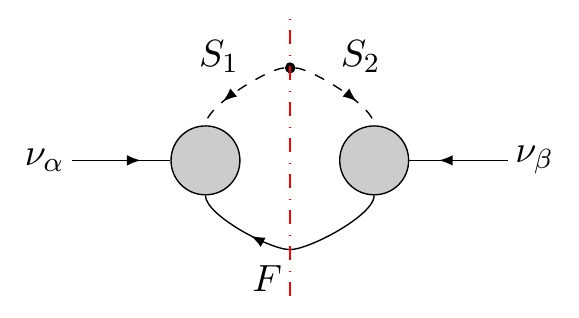}
 \caption{\label{fig:GeneralLoopDiagram2}}
\end{subfigure}
\end{center}
\caption{General examples of loops
contributing to the neutrino mass which can be disconnected by
cutting a pair of propagators.\label{fig:GeneralLoopDiagram.0}}
\end{figure}

The simplest example of this type of models
is Ma's scotogenic model~\cite{Ma:2006km}, which introduces one additional inert Higgs doublet $\eta$ as well as three RH neutrinos $N_i$. Neutrino masses are generated with the internal propagators $\langle NN^Tc\rangle$ and $\langle\eta^0\eta^0\rangle$ with $\eta^0$ being the neutral component of the inert Higgs doublet $\eta$.

The one loop suppression does not suffice to explain the smallness
of neutrino masses by itself. A further suppression is needed,
which might be due to the smallness of the lepton number violating
mass insertions that are indicated by crosses compared to the
overall masses of the particles propagating in the loops. Another
explanation of the additional suppression might be a sequence of
symmetry breaking which naturally suppresses certain couplings
(See e.g.~\cite{Farzan:2010mr}).

\subsection{Reduction of multi-loop contribution to one-loop\label{theorem}}

Let us consider a general loop contributing to the neutrino mass
which can be disconnected by cutting a pair of propagators as
shown in \Figref{fig:GeneralLoopDiagram}.
If, as shown in
\Figref{fig:GeneralLoopDiagram}, the
 fermionic propagator is of chirality flipping nature $\langle F_{1}
 F_{2}^Tc\rangle$, the vertices of the following types will be
 allowed by the $G_\nu=Z_n$ or $U(1)$ symmetry:
 \be \label{vertices} S_1 L_\alpha F_{1} \ \ \ \ {\rm and}
 \ \ \ \ S_2 L_\beta F_{2} \ . \ee
 If both the Higgs fields are attached to this pair of lines,
 these two vertices can be made $SU(2)\times U(1)$ invariant, too.
 As a result, a one-loop diagram contributing to the neutrino mass
 with the couplings in Eq. (\ref{vertices}) can be formed.

On the other hand, if the propagator is of the chirality-flipping form $\langle c
F_{1}^*F_{2}^\dagger \rangle$, the  vertices $ S_1^\dagger L_\alpha F_{1} $ and
$ S_2^\dagger  L_\beta F_{2}$ are allowed by the
$G_\nu=Z_n$ or $U(1)$ symmetry but they may violate $U(1)$
hypercharge. Thus, unlike the previous case, the
 presence of a one-loop contribution to the
neutrino mass is not guaranteed. On the other hand, if the
fermionic line is of the chirality-conserving form, {\it i.e.,} $
\langle F F^\dagger\rangle$, see \Figref{fig:GeneralLoopDiagram2},
vertices in Eq. (\ref{vertices}) might violate the $G_\nu=Z_n$ or
$U(1)$ symmetry and again, a one-loop contribution to the neutrino
mass does not necessarily exist.

  In
summary, {\it if there exists a multi-loop contribution to the
Weinberg operator, $(HL_\alpha)(HL_\beta)$ compatible with
$G_\nu=Z_n$ or $U(1)$ which can be disconnected by cutting a pair
of fermionic and scalar lines, there will be also a one-loop
contribution to the neutrino mass provided that (i) both Higgs
fields are attached to these two lines; (ii) the fermionic
propagator in question is chirality-flipping and of form $\langle
F_1 F_2^Tc \rangle$.}

\begin{figure}
\begin{center}
\begin{subfigure}{0.3\textwidth}
\FMDG{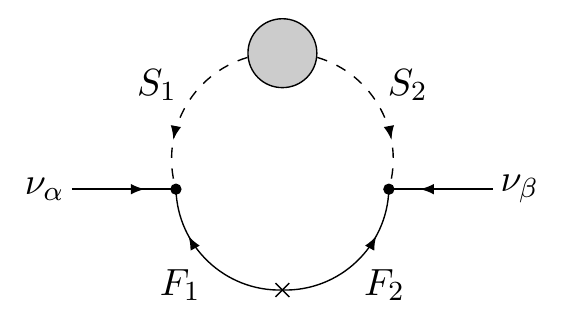}
\caption{scalar line\label{fig:WaveFunctionRenormalizationScalar}}
\end{subfigure}
\hspace{1cm}
\begin{subfigure}{0.3\textwidth}
\FMDG{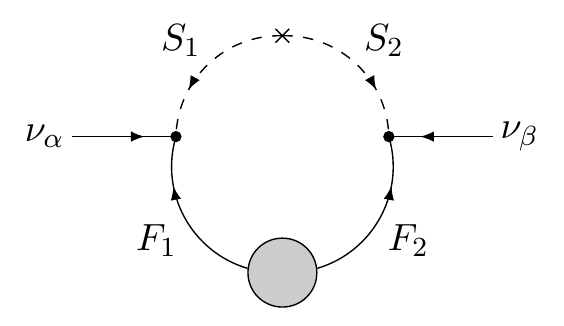}
\caption{fermion line\label{fig:WaveFunctionRenormalizationFermion}}
\end{subfigure}
 \caption{ Wavefunction renormalization of internal propagators. \label{fig:WaveFunctionRenormalization}}
\end{center}
\end{figure}

Let us now consider a general multi-loop diagram of form shown in
\Figref{fig:WaveFunctionRenormalizationScalar} in which the
internal loop only gives a correction to the wave function of the
scalar. Topologically such a diagram is distinguished from the
rest by the fact that by cutting the scalar lines directly
connected to $\nu_\alpha$ and $\nu_\beta$, this line will be
disconnected.  Figs \ref{fig:2loop.d}, \ref{fig:2loop.e}, and
\ref{fig:2loop.f} are examples of such diagrams but  Figs.
\ref{fig:2loop.a} and \ref{fig:2loop.b}  are not. If a
contribution of this type exists, the $G_\nu=Z_n$ or $U(1)$
symmetry allows a term such as $S_1S_2$, too. Depending on the
electroweak behavior of $S_1$ and $S_2$, this mass term can result
from terms such as $\epsilon_{ab} S_1^a H^b S_2$ (for singlet
$S_2$ and doublet $S_1$ with $Y_{S_1}+Y_{S_2}=-1$), $\epsilon_{ab}
S_1^aS_2^b$ (for doublets $S_1$ and $S_2$ with
$Y_{S_1}+Y_{S_2}=0$), $\epsilon_{ab}\epsilon_{cd}S_1^a H^b S_2^c
H^d$ (for doublets $S_1$ and $S_2$ with $Y_{S_1}+Y_{S_2}=-2$) or
$S_1 S_2$ (for singlets $S_1$ and $S_2$ with $Y_{S_1}+Y_{S_2}=0$).
As a result, these diagrams are always accompanied by a one-loop
diagram.

Let us now discuss  diagrams of type in
\Figref{fig:WaveFunctionRenormalizationFermion} in which the
internal loop gives correction to the wave function of the
propagating fermion. Similarly to the correction to the wave
function of the scalar propagator, the wave function correction to
the fermion propagator can also be written as a $F_1F_2$  mass
term which respects the $G_\nu=Z_n$ or $U(1)$ symmetry.  However,
if $F_1$ and $F_2$ are both electroweak doublets, $F_1F_2$ will
form an electroweak triplet. Thus, two factors of $\langle
H\rangle$ are needed to contract it to a $SU(2)\times U(1)$ singlet. In
other words, the corresponding term will be non-renormalizable.
Hence, this diagram is not necessarily accompanied by a lower loop
contribution depending on the electroweak structure of the
fermions.

Let us suppose a coupling of form $g_\alpha SFL_\alpha$ compatible with $G_\nu=Z_n$ or $U(1)$ exists. There must be another $F^\prime$ with
$\alpha_{F^\prime}=-\alpha_F$ to obtain a Dirac mass term for $F$ (either directly or after electroweak symmetry breaking).  The  $G_\nu=Z_n$ or $U(1)$ symmetry does not forbid a term of form $g_\alpha^\prime S^\dagger F^\prime L_\alpha$. The neutrinos then obtain a Majorana mass proportional to $g_\alpha g_\beta^\prime$ at one loop. In the discussion of higher loops, we implicitly assume that such a possibility is forbidden by other symmetries such as the electroweak symmetry.

%%%%%%%%%%%%%%%%%%%%%%%%%%%%%%%%%%%%%%%%%%%%%
%%%%%%%%%%%%%%%%%%%%%%%%%%%%%%%%%%%%%%%%%%%%%5
%%%%%%%%%%%%%%%%%%%%%%%%%%%%%%%%%%%%%%%%%%%%%%5
%%%%%%%%%%%%%%%%%%%%%%%%%%%%%%%%%%%%%%%%%%%%%%%%%%

\subsection{Two loop}
\begin{figure}
\begin{center}
\begin{subfigure}{0.3\textwidth}
\FMDG{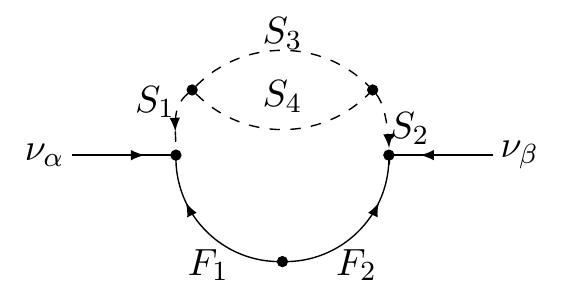}
\caption{\label{fig:2loop.d}}
\end{subfigure}
\begin{subfigure}{0.3\textwidth}
\FMDG{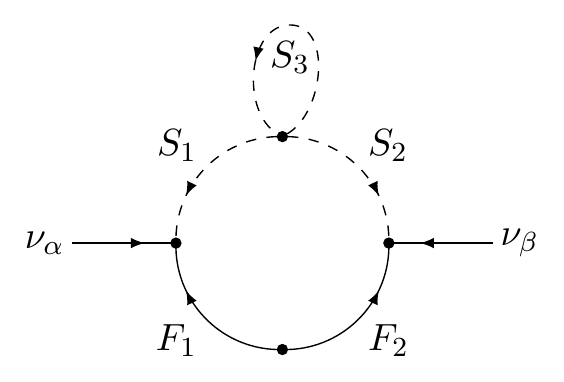}
\caption{\label{fig:2loop.e}}
\end{subfigure}
\begin{subfigure}{0.3\textwidth}
\FMDG{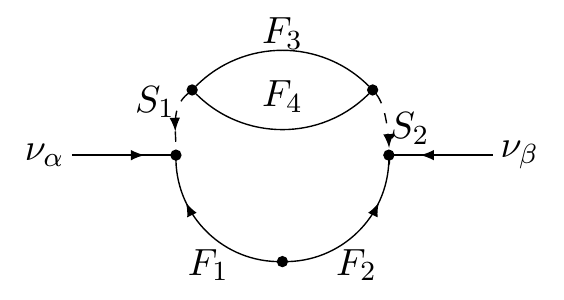}
\caption{\label{fig:2loop.f}}
\end{subfigure}
\begin{subfigure}{0.3\textwidth}
\FMDG{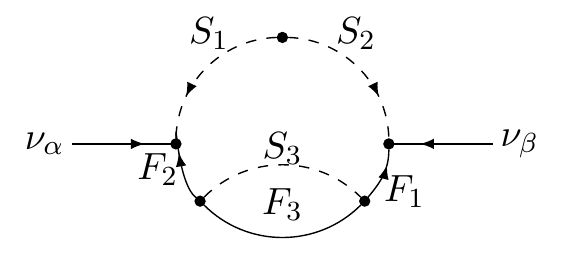}
\caption{\label{fig:2loop.c}}
\end{subfigure}
\begin{subfigure}{0.3\textwidth}
\FMDG{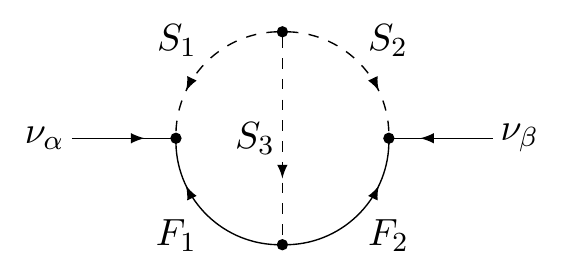}
\caption{\label{fig:2loop.a}}
\end{subfigure}
\begin{subfigure}{0.3\textwidth}
\FMDG{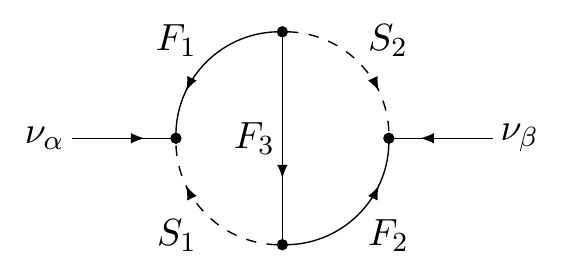}
\caption{\label{fig:2loop.b}}
\end{subfigure}
\caption{Two loop diagrams\label{fig:2loop}}
\end{center}
\end{figure}

%%%%
At two-loop level, there are more possible diagrams. In the
following, we again take $G_\nu=Z_n~{\rm or}~U(1)$ and discuss in
which cases the symmetries forbid the lower loop contribution to
the neutrino mass. In \Figref{fig:2loop}, for the sake of
simplicity, some of the scalar lines are marked by a single letter
such as $S_1$ and $S_2$. If the external Higgs is attached to any
of these lines, this line will in fact involve more than a single
field. When the fermionic propagator involves an even number of
fields, the chirality will be flipped because it requires a mass
term such as $F_1F_2$.  In the diagrams, the arrow indicates the
direction of the flow of the $G_\nu$ charge, both for fermion
propagator and scalar propagator. Notice that for the fermion
lines, the chirality might flip but the direction of the arrow
will remain the same. If a chirality flip is required ({\it e.g.},
\Figref{fig:2loop.b}), the fermionic line will involve the
fermion, $F$ and its partner $F^\prime$ with opposite $G_\nu$
charge that together form a Dirac mass. This simplified way of
marking does not generally affect our discussion below. We will be
more specific when it does.

As discussed in the previous section,
the  diagrams \ref{fig:2loop.d}, \ref{fig:2loop.e}, and \ref{fig:2loop.f} are always accompanied by a one-loop
 diagram but  this is not necessarily the case for the diagram in \Figref{fig:2loop.c}, as it has been discussed in the previous section.
The diagrams \ref{fig:2loop.a} and \ref{fig:2loop.b} cannot be reduced to a one loop diagram,
as long as $S_3$ and $F_3$ transform non-trivially under
$G_\nu=Z_n~{\rm or}~U(1)$, so they
can give the dominant contribution to neutrino mass.

%%%%%%%%%%%%%%%%%%%%%%%%%%%%%%
\subsection{Three loop}
\begin{figure}[tb]
\begin{center}
\begin{subfigure}{0.3\textwidth}
\FMDG{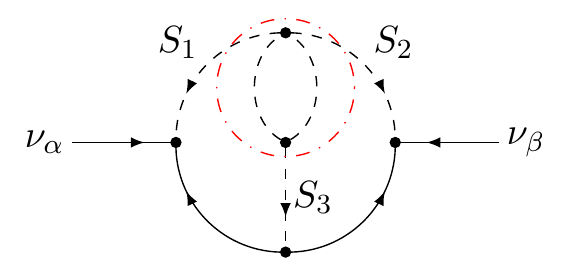}
\caption{\label{fig:3loopPlanar.a}}
\end{subfigure}
\begin{subfigure}{0.3\textwidth}
\FMDG{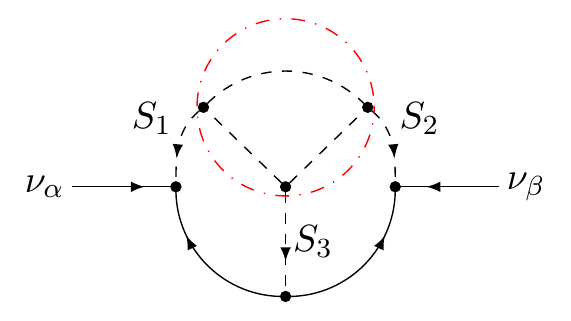}
\caption{\label{fig:3loopPlanar.b}}
\end{subfigure}
\begin{subfigure}{0.3\textwidth}
\FMDG{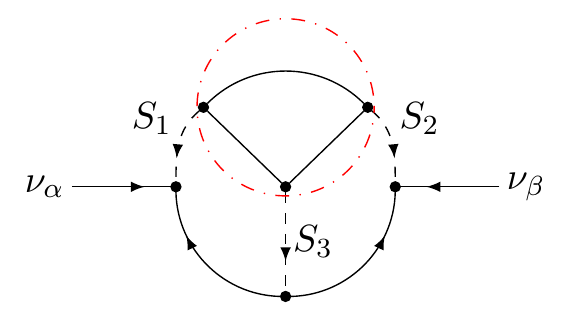}
\caption{\label{fig:3loopPlanar.bf}}
\end{subfigure}
\begin{subfigure}{0.3\textwidth}
\FMDG{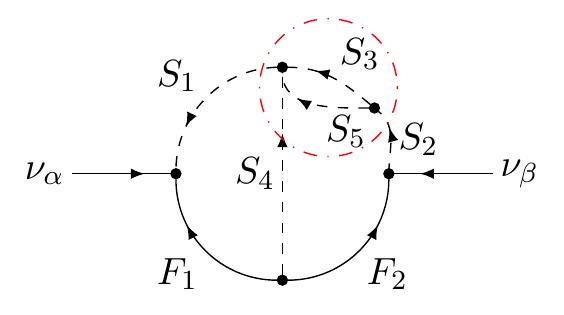}
\caption{\label{fig:3loopPlanar.Kb}}
\end{subfigure}

\caption{Planar three loop diagrams.
\label{fig:3loopPlanar.1}}
\end{center}
\end{figure}
\begin{figure}[tb]
\begin{center}
\begin{subfigure}{0.3\textwidth}
\FMDG{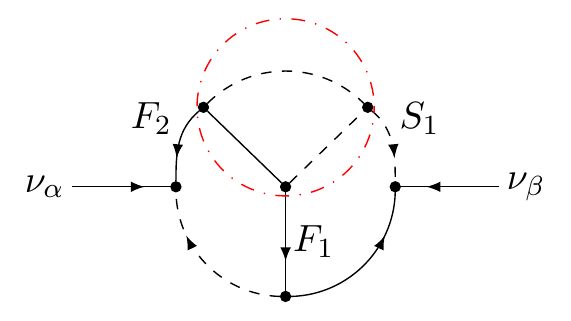}
\caption{\label{fig:3loopPlanar.d}}
\end{subfigure}
\begin{subfigure}{0.3\textwidth}
\FMDG{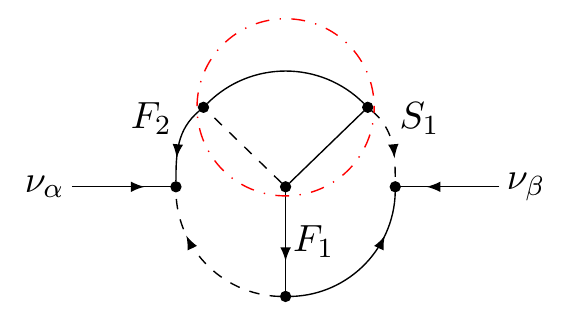}
\caption{\label{fig:3loopPlanar.e}}
\end{subfigure}
\begin{subfigure}{0.3\textwidth}
\FMDG{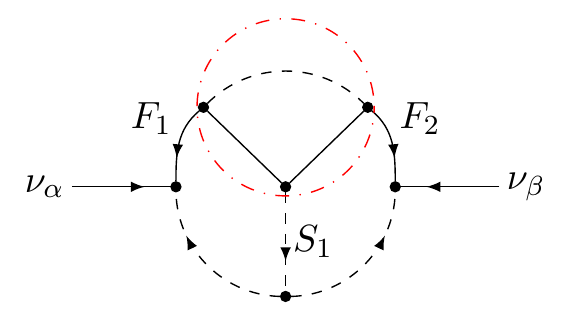}
\caption{\label{fig:3loopPlanar.f}}
\end{subfigure}
\begin{subfigure}{0.3\textwidth}
\FMDG{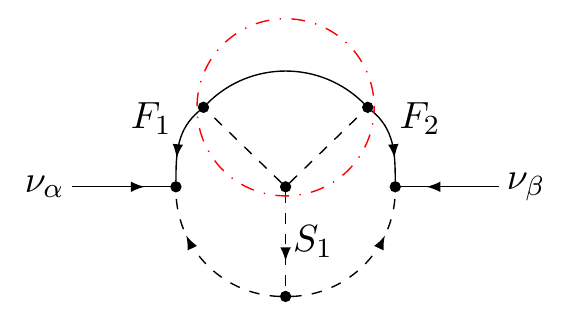}
\caption{\label{fig:3loopPlanar.g}}
\end{subfigure}
\caption{Planar three loop diagrams.
\label{fig:3loopPlanar.2}}
\end{center}
\end{figure}
\begin{figure}[tb]
\begin{center}
\begin{subfigure}{0.3\textwidth}
\FMDG{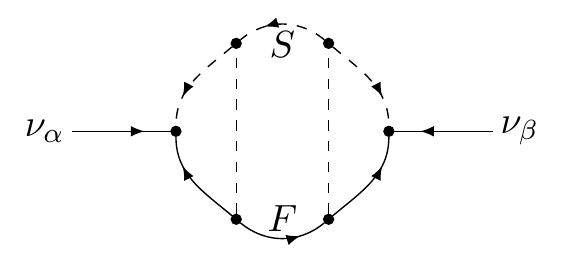}
\caption{\label{fig:3loopPlanar.h}}
\end{subfigure}
\begin{subfigure}{0.3\textwidth}
\FMDG{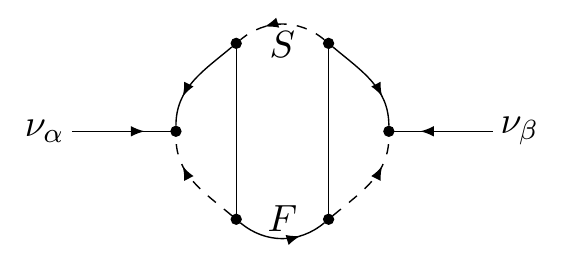}
\caption{\label{fig:3loopPlanar.i}}
\end{subfigure}
\begin{subfigure}{0.3\textwidth}
\FMDG{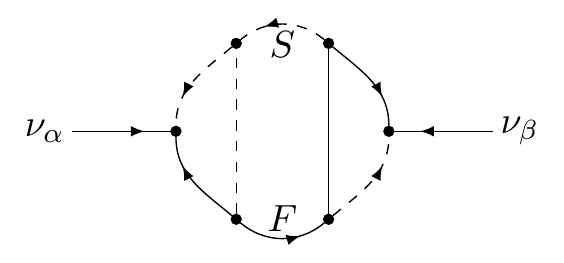}
\caption{\label{fig:3loopPlanar.j}}
\end{subfigure}
\caption{Planar three loop diagrams.
\label{fig:3loopPlanar.3}}
\end{center}
\end{figure}
\begin{figure}[tb]
\begin{center}
\resizebox{0.3\linewidth}{!}{\FMDG{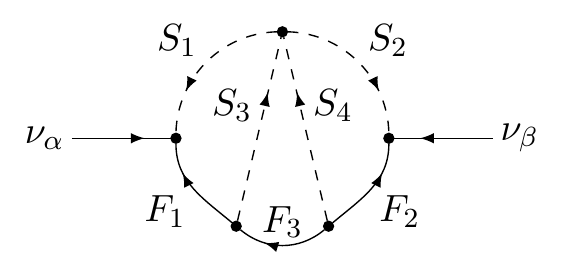}}
\caption{Planar three loop diagrams.
\label{fig:3loopPlanar.4}}
\end{center}
\end{figure}

%%%%%%%%%%%%%%%%%%%%%%%%%%%%%%%

\begin{figure}[tb]
\begin{center}
\begin{subfigure}{\textwidth}
\centering
\FMDG{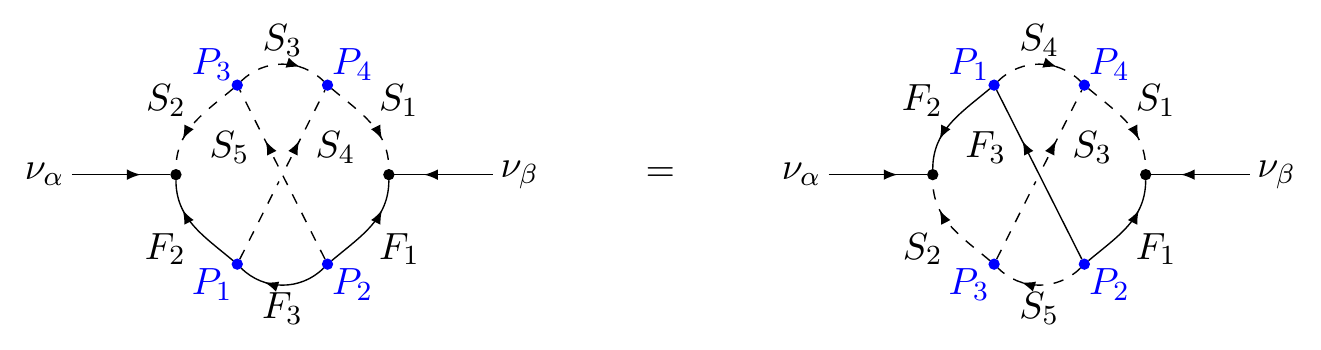}
\caption{\label{fig:3loopNonPlanar.a}}
\end{subfigure}
\begin{subfigure}{\textwidth}
\centering
\FMDG{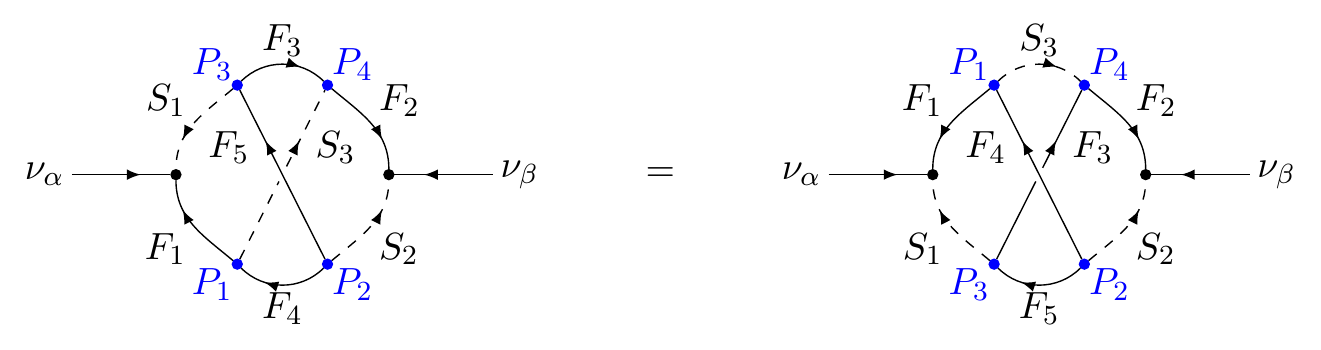}
\caption{\label{fig:3loopNonPlanar.b}}
\end{subfigure}
\caption{Non-planar three loop diagrams} \label{fig:3loopNonPlanar}
\end{center}
\end{figure}

Before starting the discussion of the three-loop diagrams, we
emphasize that the comment in the first paragraph of the previous
section on marking the propagators applies here, too.
 The  three-loop diagrams contributing to the neutrino mass can be
divided into three categories: (i) diagrams in which the inner
loops correct the wave function of the internal lines. Such
diagrams are already discussed in sect.~\ref{theorem}. (ii) Planar
diagrams shown in Figs.~\ref{fig:3loopPlanar.1},
\ref{fig:3loopPlanar.2}, \ref{fig:3loopPlanar.3} and
\ref{fig:3loopPlanar.4}. We shall discuss all these diagrams in
detail below in context of $G_\nu=Z_n$ and $G_\nu=U(1)$. (iii)
Non-planar diagrams shown in \Figref{fig:3loopNonPlanar} which
will be discussed in detail in this section in context of
$G_\nu=Z_n$ and briefly for $G_\nu=U(1)$. Although at first sight
it seems there are four non-planar three-loop diagrams
contributing to the neutrino mass, only two are independent. As
demonstrated in \Figref{fig:3loopNonPlanar} twisting the vertices
denoted by $P_1$ and $P_3$, the diagrams on the left- and
right-hand sides of \Figref{fig:3loopNonPlanar} convert into each
other. There are therefore only two distinct non-planar diagrams.

The $G_\nu=Z_n$ or $G_\nu=U(1)$ symmetries  do not forbid lower
order loop contributions for any of the planar diagrams besides
the one in \Figref{fig:3loopPlanar.4}, but the pattern of
electroweak symmetry breaking as well as the requirement of
chirality flipping might prevent some. Let us discuss this
possibility in detail. In \Figref{fig:3loopPlanar.1} and
\Figref{fig:3loopPlanar.2}, the internal loop in the red dashed dotted
circle can be replaced by an effective $G_\nu$ conserving vertex.
Let us first consider the diagrams in \Figref{fig:3loopPlanar.1}.
 For a  Lagrangian symmetric  under $G_\nu=Z_n$ (or under $G_\nu=U(1)$),
  the presence of these diagrams is possible only if
$\alpha_{S_i}$ associated with $S_1$, $S_2$ and $S_3$ add up to an
integer times $n$ (or add up to zero).
 Thus, the $G_\nu$
invariant Lagrangian contains vertices of type $S_1 S_2 S_3+\hc$ unless
it is forbidden by some other symmetry. In particular let's consider the
$SU(2) \times U(1)$ symmetry: each line
might involve more than one field with different $SU(2)\times U(1)$
quantum numbers and we
should specify the fields that directly leave the red dashed dotted circle.
In case that $Y_{S_1}+Y_{S_2}+Y_{S_3}=0$ or $-1$, the
corresponding vertex can be just $S_1 S_2 S_3+\hc$ or
$S_1S_2S_3\braket{H}+\hc$, respectively. However, there is no renormalizable vertex
of form $S_1S_2S_3$ if $Y_{S_1}+Y_{S_2}+Y_{S_3}<-1$
 and no corresponding two-loop diagram.
 Similarly, in case of the diagrams in \Figref{fig:3loopPlanar.2},
 the loop in the dashed dotted red circle can be also replaced  by
a renormalizable  $Z_n$ invariant vertex
 of form $S_1^\dagger  F_1^\dagger F_2^\dagger$ unless
 $Y_{S_1}+Y_{F_1}+Y_{F_2}\ne 0$. Notice that we implicitly assume
 that both left-handed fields $F_1$ and $F_2$ leave the red dashed dotted circle;
 {\it i.e.,} the corresponding effective vertex is of form
 $S_1^\dagger F_1^\dagger F_2^\dagger$ rather than $S_1^\dagger F_2^\dagger F_1$ or  $S_1^\dagger F_1^\dagger F_2$
 which are  forbidden by Lorentz structure.
Let us consider the loop on the right-hand side in the diagrams of \Figref{fig:3loopPlanar.3}. If, as indicated in the
figures, the fermion marked with $F$ enters this loop, the loop can
 be replaced
 by a renormalizable $Z_n$ invariant Yukawa coupling of form
 $S^* F L_\beta$, which conserves hypercharge if $Y_S=Y_F  -1$.

The implication of $Z_n$ for \Figref{fig:3loopPlanar.4}  as well as the non-planar diagrams in \Figref{fig:3loopNonPlanar} is more
complicated. In particular, $Z_n$ does not always allow them to be
accompanied by a dominating two loop contribution to the neutrino mass.
Let us first consider diagram \ref{fig:3loopPlanar.4}.
The $Z_n$ symmetry implies
\begin{align}\label{set0}
\alpha_{F_1}&=-\alpha_{S_1}+n k_1 &
\alpha_{F_3}&=\alpha_{S_3}-\alpha_{S_1}+nk_3\\\nonumber
\alpha_{F_2}&=-\alpha_{S_2}+n k_2&
\alpha_{S_4}&=\alpha_{S_1}+\alpha_{S_2}-\alpha_{S_3}+nk_4
\end{align}
where $k_i$ are arbitrary integers. We are interested to find out
whether there is a $Z_n$ symmetry with certain $\alpha_\phi$
assignment which is compatible with the diagram in
\Figref{fig:3loopPlanar.4}, {\it i.e.}, satisfies
Eqs.~(\ref{set0}) but forbids lower loop contributions. To answer
this question, we have solved  equations (\ref{set0}) under the
condition that none of the one- and two-loop diagrams respectively
in \Figref{fig:OneLoop} and \Figref{fig:2loop}, is allowed by the
$Z_n$ symmetry. The values of $\alpha_{S_i}$ and $\alpha_{F_j}$
can be set such that the $Z_n$ symmetry forbids the lower orders
of contributions to the neutrino mass for $n\ge16$. One example is
the charge assignment
\begin{align}
\alpha_{S_1}&=1&
\alpha_{S_2}&=3&
\alpha_{S_3}&=9&
\alpha_{S_4}&=11&
\alpha_{F_1}&=15&
\alpha_{F_2}&=13&
\alpha_{F_3}&=8
\end{align}
for $Z_{16}$.
However, for smaller values of $n$ such an assignment is not
possible.

Similarly in case of the non-planar diagrams,  for certain assignments of $\alpha_{S_i}$ and $\alpha_{F_j}$, the $Z_n$ symmetry forbids all diagrams with a lower loop order. Let us consider diagram \Figref{fig:3loopNonPlanar.a}.
 The $Z_n$ symmetry leads
to the following relations for the different fields propagating
inside the loops:
\begin{align}\label{set1}
\alpha_{F_1}&=-\alpha_{S_1}+nk_1&
\alpha_{S_3}&=\alpha_{S_5}-\alpha_{S_2}+nk_4\\\nonumber
\alpha_{F_2}&=-\alpha_{S_2}+nk_2&
\alpha_{S_4}&=\alpha_{S_1}+\alpha_{S_2}-\alpha_{S_5}+nk_5\\\nonumber
\alpha_{F_3}&=\alpha_{S_1}-\alpha_{S_5}+nk_3
\;. \end{align}
We have found that
the smallest value of $n$ for which the $Z_n$ symmetry forbids
lower order loop contribution to the neutrino mass is $n=16$. One particular example is
\begin{align}
\alpha_{S_1}&=2&
\alpha_{S_2}&=6&
\alpha_{S_3}&=13&
\alpha_{S_4}&=5&
\alpha_{S_5}&=3&
\alpha_{F_1}&=14&
\alpha_{F_2}&=10&
\alpha_{F_3}&=15
\end{align}
for $Z_{16}$.
That is for $n<16$ any possible assignment of $\alpha_{S_i}$
and $\alpha_{F_i}$ will also lead to a dominant lower loop diagram. In this analysis, only the $Z_n$
symmetry is considered. Of course, depending on the field content
of the specific model, the pattern of the hypercharge breaking
as well as the form of chirality flipping might also
forbid the lower loop contribution.

A similar analysis is performed for the diagram in \Figref{fig:3loopNonPlanar.b}.
To be compatible with the $Z_n$ symmetry, the following set of equations has to be satisfied
\begin{align}\label{set2}
\alpha_{F_1}&=-\alpha_{S_1}+ nk_1&
\alpha_{F_4}&=-\alpha_{S_1}+\alpha_{S_3}+nk_4\\\nonumber
\alpha_{F_2}&= -\alpha_{S_2}+nk_2&
\alpha_{F_5}&=\alpha_{S_1}-\alpha_{S_2}-\alpha_{S_3}+nk_5\\\nonumber
\alpha_{F_3}&=-\alpha_{S_2}-\alpha_{S_3}+nk_3
\end{align}
for any integers $k_i$.  Here, $n=16$ is again the smallest value
of $n$ for which $Z_n$
 forbids a lower loop contribution to the neutrino
mass. One example of  charge assignment for $Z_{16}$
 forbidding
lower loop contributions  is
\begin{align}
\alpha_{S_1}&=2&
\alpha_{S_2}&=6&
\alpha_{S_3}&=3&
\alpha_{F_1}&=14&
\alpha_{F_2}&=10&
\alpha_{F_3}&=7&
\alpha_{F_4}&=1&
\alpha_{F_5}&=9\;.
\end{align}
Notice that all three of the three loop diagrams require $n\geqslant
{16}$ to forbid  lower order loop diagrams. We will first show
that the conditions for \Figref{fig:3loopNonPlanar.a} and
\Figref{fig:3loopNonPlanar.b} are equivalent. If we identify the
charges in \Eqref{set1}, as indicated in \Tabref{tab:transFigs10},
it is straightforward to show, that we obtain a set of equations,
which is equivalent to \Eqref{set2} and vice versa. Note that we
are identifying the charges of the fermions in one diagram with
the ones of the scalars in the other diagram and vice versa.
\begin{table}
\begin{center}
\begin{tabular}{l|cccccccc}
\Figref{fig:3loopNonPlanar.a} & $S_1$ & $S_2$  & $S_3$ & $S_4$& $S_5$ & $F_1$& $F_2$  & $F_3$ \\\hline
\Figref{fig:3loopNonPlanar.b} & $F_2$ &  $F_1$ & $F_5$ & $F_4$ & $F_3$ & $S_2$& $S_1$  & $S_3$
\end{tabular}
\end{center}
\caption{Translation table between \Figref{fig:3loopNonPlanar.a} and \Figref{fig:3loopNonPlanar.b}.\label{tab:transFigs10}}
\end{table}
Remember that  to form a Dirac mass term for the new fermions, we have to introduce a
partner for each fermion so the charges of scalars
 and fermions may be  treated on an equal footing. As a result,
 the equivalence of the
conditions for the absence of  lower loop contributions directly
follows. The equivalence of charge assignments also
implies that for a
 given $n\geqslant{16}$, there must be exactly the same number of
 possible charge configurations for each diagram that forbids the
 lower order loops. In fact, solving the equations for $n=16$, we find that there
 are exactly $32$ possible charge configurations for each of the
 diagrams \Figref{fig:3loopNonPlanar.a} and
 \Figref{fig:3loopNonPlanar.b} that forbid lower order contributions.

It is straightforward to verify that after replacing \begin{align}
\alpha_{S_3}&\leftrightarrow \alpha_{S_5} &
\alpha_{F_3}&\leftrightarrow -\alpha_{F_3}\;,
\end{align}
Eqs.~(\ref{set0}) will be equivalent to the first four equations
in Eq.~(\ref{set1}). As a result, if for a given $n$ there is an
assignment of charges for the field content of
\Figref{fig:3loopNonPlanar.a} that forbids lower loop
contributions to the neutrino mass, the corresponding assignment
for  \Figref{fig:3loopPlanar.4} will also forbid lower order
contributions. However, the opposite statement is not valid.  In
fact, the presence of an extra scalar field in case of
\Figref{fig:3loopNonPlanar.a} gives more freedom to construct more
possible lower loop diagrams. Setting $n=16$, we find 112
solutions for \Figref{fig:3loopPlanar.4} that forbid lower loop
contributions but only 32 such solutions for diagrams in
\Figref{fig:3loopNonPlanar.a}.

In this discussion, we have assumed that each fermionic (scalar)
propagator has an independent $\alpha_{F_i}$ ($\alpha_{S_j}$)
value. The relation between $\alpha_{F_i}$ and $\alpha_{S_j}$
comes from the requirement that this diagram respects $Z_n$.
However, within specific models, there might be more restrictions.
For example, let us assume that $S_1$ and $S_2$ in diagram
\Figref{fig:3loopNonPlanar.a} are the same fields. We then
conclude that for any value of $n$ and any choice of
$\alpha_{F_i}$ and $\alpha_{S_j}$ (provided that
$\alpha_{S_1}=\alpha_{S_2}$), the $Z_n$ symmetry allows the
diagram in \Figref{fig:3loopNonPlanar.a} to be accompanied with
two diagrams of form in \Figref{fig:2loop.a} with the following
replacements of the fields in \Figref{fig:2loop.a}:
$(S_1,\,S_2,\,S_3,\,F_1,\,F_2)\to (S_1,\,S_3,\,S_5^*,\,F_1,\,F_3)$
or
$(S_1,\,S_2,\,S_3,\,F_1,\,F_2)\to(S_1,\,S_3^*,\,S_4^*,\,F_1,\,F_3^\prime)$
where $F_3^\prime$ is the Weyl fermion with charge opposite to
that of $F_3$ that together form a Dirac mass term $F_3
F_3^\prime$.
Depending
on the field content of the model, it is possible that one or both
of these two-loop diagrams are forbidden by $SU(2) \times U(1)$ or
by pattern of chirality flipping. For example, if in
\Figref{fig:3loopNonPlanar.a}, one of the Higgs fields is attached
to $S_4$ and the other to $S_5$, neither of these two-loop
diagrams can exist because each violates hypercharge only by one
unit.

Let us now briefly discuss the possibility of replacing $Z_n$ with
a continuous $U(1)$. In this case setting $n=0$, relations in Eqs.
(\ref{set1}) and (\ref{set2}) remain valid. For a $U(1)$ with
general $\alpha_{S_i}$ and $\alpha_{F_j}$, the three-loop
non-planar diagrams and the planar diagram in
\Figref{fig:3loopPlanar.4} cannot be accompanied with a lower loop
contribution unless in very specific cases. For example, in the
specific case that $\alpha_{S_4}= -\alpha_{S_5}$, there might be
also a one-loop
 contribution to the neutrino
mass accompanying the diagram in \Figref{fig:3loopNonPlanar.a}.

%%%%%%%%%%%%%%%%%%%%%%%%%%%%%%%%%%%%%%%%%%%%%%%%%%%%%%%%%%%%%%%%%%%%

\section{General symmetry\label{sec:generalisation}}
In previous sections, we focused on the implications of
$G_\nu=Z_n$ or $U(1)$ symmetry on the neutrino mass generation at
loop levels. As we shall discuss below, some of these discussions
can be applied for a general symmetry group $G_\nu$. We will make
a further generalization in this section. Motivated by the DM
models, in the previous sections, we have assumed that the SM
particles are all invariant under the $G_\nu$ symmetry. In this
section, we discuss the consequences of relaxing this assumption.
We will still assume that none of the scalar fields which
non-trivially  transform under $G_\nu$ ({\it i.e.,} are not
invariant under $G_\nu$) receives a VEV. Thus, the SM Higgs is
invariant under $G_\nu$.

 Let us reconsider
coupling (\ref{L-Y}) and review the results that we found in the
previous sections. \begin{itemize}
\item As long as all new neutral fermions non-trivially transform under $G_\nu$, the Dirac mass
term for neutrinos will be forbidden by this symmetry to all
orders in perturbation theory.
\item Consider a general $n$-loop diagram
contributing to neutrino mass. Suppose that there is a sub-diagram
within this diagram that absorbs scalar lines $S_1$, $S_2$ and
$S_3$. The existence of such a sub-diagram implies that a combination
of form $S_1S_2S_3$ is invariant under $G_\nu$. If no Higgs VEV is
attached to this sub-diagram, it means this combination is also
invariant under electroweak symmetry and this term is allowed in the Lagrangian.
Thus, there should be a lower order contribution to the neutrino mass
where this sub-diagram is replaced by the $S_1S_2S_3$ vertex.
Similarly, if there is only one Higgs VEV attached to this
sub-diagram, a renormalizable coupling of form $S_1 S_2 S_3 H$
exists in the Lagrangian which can lead to a lower order
contribution to the neutrino mass. However, if both Higgs VEVs of
the Weinberg operator are attached to the sub-diagram, the
corresponding electroweak vertex will not be renormalizable.
Similarly, there might be a sub-diagram in which two fermions
($F_1$ and $F_2$) and one scalar ($S_1$) leave the sub-diagram.
Examples are shown in \Figref{fig:3loopPlanar.2}, where the sub-diagram is inside
the red circle. Regardless of the details of $G_\nu$ and whether
the SM particles are invariant under it or not, a vertex of form
$F_1 F_2 S_1$ is  invariant under $G_\nu$. If it is also invariant
under electroweak symmetry ({\it i.e.,} if no Higgs  VEV is
attached to the sub-diagram) this renormalizable term will be
present in the Lagrangian. This means this diagram is accompanied
by a lower order diagram in which the sub-diagram is replaced by
vertex of form $F_1 F_2 S_1$.
\item Let us now consider a general diagram that contains a
sub-diagram which is a correction to the self-energy of a scalar
line, that is the external lines that are attached to this
sub-diagram are two scalars $S_1$ and $S_2$ (See \Figref{fig:WaveFunctionRenormalizationScalar}).
 The $S_1S_2$ term
will be $G_\nu$ invariant and in case that less than three Higgs
VEVs are attached to this sub-diagram, it can be made electroweak
invariant by adding appropriate number of Higgs fields ({\it
i.e.,} $S_1S_2$, $S_1 S_2 H$ or $S_1S_2 H H$). The diagram is
accompanied by another one in which the sub-diagram is replaced by
the corresponding renormalizable vertex. Similar consideration
holds for a sub-diagram giving correction to the fermion self
energy but in this case  more than one Higgs fields  cannot  be
added otherwise the corresponding term will be non-renormalizable.
Notice that this consideration holds valid regardless of  the
details of the $G_\nu$ symmetry and the behavior of the SM
fermions under this symmetry.
\item The theorem of section \ref{theorem},
regarding diagrams of form shown in \Figref{fig:GeneralLoopDiagram.0} holds valid for a
general $G_\nu$ independent of the behavior of the SM fermion
under the symmetry transformation.
\item As shown for the special case of $G_\nu=Z_n$, the $G_\nu$
symmetry can forbid lower order contribution for the two-loop
diagram of topology shown in \Figref{fig:2loop.a} and
\Figref{fig:2loop.b}, as well as the three-loop diagrams of
topology in Figs. \ref{fig:3loopPlanar.4} and
\ref{fig:3loopNonPlanar}.
\end{itemize}

%%%%%%%%%%%%%%%%%%%%%%%%%%%%%%%%%%%%%%%%%%%%%%%%%%%%%%%%%%%%%%%%%%%%

\section{Lepton flavor violation\label{sec:lfv}}
\begin{figure}
\begin{center}
\resizebox{0.3\linewidth}{!}{\FMDG{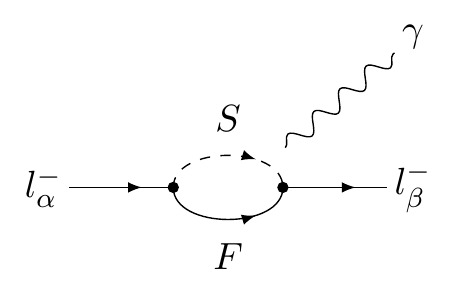}}
\end{center}
\caption{LFV diagram. The photon is emitted either from the
initial or
  final state or from the charged particle in the loop.\label{fig:LFVOneLoop}}
\end{figure}

After discussing the general structure of the different diagrams
leading to neutrino masses, we will discuss lepton flavor
violating rare decays which have proven to lead to strong
constraints on radiative neutrino mass models. The general Yukawa
coupling in Eq.~(\ref{L-Y}) includes LFV couplings of left-handed
charged leptons, $l^-_\alpha$, $g_{ij \alpha} s_i f_j l^-_\alpha$
where $s_i$ and $f_j$ are components of the multiplets $S_i$ and
$F_j$ such that the following relation holds
 valid between their
electric charges:
$$ Q_{s_i}+Q_{f_j}=-Q_{l_\alpha^-}=1\ .$$
This LFV coupling will lead to LFV rare decay $l_\alpha^- \to
l_\beta^- +\gamma$ as shown in \Figref{fig:LFVOneLoop}. Note that
in contrast to the case of contributions to the neutrino masses,
the LFV rare decays are generically allowed at one loop order
unless there is a flavor symmetry forbidding the one loop
contribution; see {\it e.g.}~\cite{Ma:2010gs}. Neglecting the
corrections of order of $(m_{l_\beta}/m_{l_\alpha})^2$, the decay
rate is \cite{Lavoura:2003xp} \be \Gamma (l_\alpha^- \to l_\beta^-
+\gamma)=\frac{\alpha_{QED} m_{l_\alpha^-}^5}{(384
\pi^2)^2}\left|\sum_{ij}\frac{g_{ij \alpha} g_{ij
\beta}^*}{m_{s_i}^2} \left( Q_{f_j}
J\left[\frac{m_{f_j}^2}{m_{s_i}^2}\right]
+I\left[\frac{m_{f_j}^2}{m_{s_i}^2}\right]\right)\right|^2 \ ,\ee
where \be I[t]=\frac{2 t^2+5t-1}{(t-1)^3}-\frac{6t^2\ln
t}{(t-1)^4} \ee and \be J[t]=\frac{3t+3}{(t-1)^2}-\frac{6t\ln
t}{(t-1)^3}\ . \ee   $I(t)$ and $J(t)$ are  monotonously
decreasing functions with the following values:
\begin{subequations}
\begin{align}
I(0)&=1& I(1)&=1/2&
I(t)&\stackrel{t\to\infty}{\longrightarrow}\frac{2}{t}\\
J(0)&=3& J(1)&=1&
J(t)&\stackrel{t\to\infty}{\longrightarrow}\frac{3}{t} \;.
\end{align}
\end{subequations}
Notice that this relation holds regardless of the representation
of the electroweak symmetry to which the new particles belong.
Larger representations can lead to several different possible loop
diagrams.

In the case that the neutrino mass originates from the three-loop
contribution, the coupling should be  of order one for
$m_{NEW}\sim 100$~GeV to account for $m_\nu \sim \sqrt{\Delta
m_{atm}^2}$. This will lead to Br($\mu \to e \gamma$) exceeding
the present bound. However, there are ways to avoid these bounds making use of a particular flavor structure.
In the following, we explain a simple and natural solution. To
reproduce two nonzero neutrino mass eigenvalues, more than one
pair of $(S,F)$ coupled to $L$ is required, which we will call
$(S_1,F_1)$ and $(S_2,F_2)$.
 Let us suppose that
$F_1$ only couples to $L_e$ while $F_2$ couples to $L_\mu$ and
$L_\tau$.  That is $g_{11\mu}=g_{11\tau}=g_{22e}=0$. In this case,
the couplings conserve $L_e$ so $\mu \to e \gamma$ will be absent.
As it is well
know~\cite{Buchmuller:1998zf,*Vissani:1998xg,*Barbieri:1998mq},
the conservation of $L_e$ leads to a vanishing first row and
column of the neutrino mass matrix, {\it i.e.,} $m_{e\alpha}=0$
for $\alpha=e,\,\mu,\,\tau$, and  can therefore only serve as a
leading order approximation of neutrino masses. If $L_e$ is softly
broken by trilinear scalar couplings,
 the vanishing elements can be
reproduced and the observed neutrino mass
 pattern can be
 reconstructed. Obviously, the breaking introduces $\mu \to e
\gamma$, but it is controlled by the smallness of
the symmetry breaking and will not be dangerous.

The new coupling can lead to a new contribution to
anomalous magnetic moment of muon as follows
$$\frac{(g-2)_\mu}{2}=\frac{m_\mu^2}{192 \pi^2} \sum_{ij}\frac{|g_{ij \mu}|^2}{m_{s_i}^2} \left( Q_{f_j} J\left[\frac{m_{f_j}^2}{m_{s_i}^2}\right]
+I\left[\frac{m_{f_j}^2}{m_{s_i}^2}\right]\right)\ .$$
Notice that for
$m_S,m_F\sim 100$~GeV and $g\sim 0.5$, this contribution can
explain the observed anomaly.

\section{Dark matter\label{sec:DM}}

The symmetry required to suppress neutrino masses  can have important implications for dark matter. 
Since we have taken the SM particles to be invariant under the $G_\nu$
symmetry, this symmetry protects the lightest new particle from
decay  and the latter constitutes a potential DM candidate. Depending on the exact form of the symmetry, there might be several stable particles and multiple DM components. We briefly discuss the main features of these models. For a detailed discussion of specific realizations, we refer the reader to Ref.~\cite{Batell:2010bp} in the case of an abelian $Z_n$ symmetry and Ref.~\cite{Adulpravitchai:2011ei} for an explicit construction of a $D_3$ model.
 We will briefly summarize the main points. In the
case of direct product groups, {\it i.e.} groups $G$ which can be
written as $G_1\times G_2$ with two arbitrary groups $G_i$, there can be two DM candidates, given by the two lightest
particles transforming non-trivially under each of the two group factors.
This happens for example for a model based on the Abelian finite
group $Z_6\cong Z_3\times Z_2$ containing two fields with $Z_6$
charges equal to $+2$ and $+3$, respectively. More generally, for
every subgroup $H$ of $G$, the lightest particle transforming
non-trivially under the subgroup might potentially be stable and a
DM candidate. This may  lead to a plethora of DM
candidates. In the case of finite abelian groups, there is a
complete classification in terms of direct products of
$Z_{p_i^{n_i}}$ factors with $p_i$ being prime numbers and $n_i$ natural
numbers. Each factor of order $p^n$ has non-trivial subgroups
$Z_{p^m}$ with $0<m<n$. Hence, there are potentially $\sum_i n_i$
DM candidates, one for each non-trivial subgroup, depending on the mass spectrum. As an example, let us consider a model with
$Z_4=Z_{2^2}$ symmetry containing two fields which under $Z_4$
transform as follows:
 $\phi_1\to e^{i\pi/2}\phi_1$, $\phi_2\to
e^{i\pi}\phi_2$. The $Z_4$
symmetry contains a $Z_2$ subgroup under which $\phi_1\to
-\phi_1$ and $\phi_2\to \phi_2$. For $m_{\phi_1}<m_{\phi_2}/2$, $\phi_1$ is the only DM candidate, since $\phi_2$ can decay into $\phi_1\phi_1$ via the coupling $\phi_1\phi_1\phi_2$. In case $m_{\phi_2}/2<m_{\phi_1}$, both fields $\phi_i$ will be stable and
therefore DM candidates. 

Generally, in presence of multiple DM candidates coannihilation will take place. However, in the models considered, particles belonging to different factors of a direct product group will not coannihilate. In the following, we will restrict ourselves to the simplest case of an abelian group, which can be written in terms of a direct product of groups without a proper subgroup, i.e. there is no coannihilation between the DM candidates.
In order to prevent having a charged DM candidate,
 the stable particles must be either neutral or in case that they are
 charged, their annihilation cross section should be much larger than $10^{-36}  \ {\rm cm}^{2}$.
We focus on the SM gauge, Yukawa interactions as well as annihilation via the Higgs portal, considering each possibility one by one. In principle, there might be a new gauge interaction contributing to DM annihilation but we will not discuss this additional extension of the models. We will mainly consider the cases in which the DM belongs to a doublet, is a singlet or combination of the two.

\mathversion{bold}
\subsection*{Annihilation via $Z$ boson exchange}
\mathversion{normal}
First, let us consider the case in which the dark matter is the neutral component of a scalar doublet of SU(2), $S$. The annihilation mode
 of $S \bar{S}$ through $s$-channel  $Z$ boson exchange is allowed with cross section given by
\begin{equation}
\langle \sigma(S \bar{S} \to f \bar{f}) v\rangle = \frac{N_c G_F^2 }{
 2 \pi}\frac{32}{ 3}\frac{(|a_L|^2+|a_R|^2) (m_S v)^2}{(1-4m_S^2/m_Z^2)^2}\;,
\end{equation} 
where  $v$ is the velocity of DM, $N_c=3(1)$ for quarks (leptons)
and  $a_L$ ($a_R$) is the coupling of the left-handed (right-handed) fermions  to the $Z$
boson. 
However, the annihilation cross section of dark matter via $Z$ bosons is directly related to the direct detection cross section. In fact, the annihilation of a complex scalar via $s$-channel $Z$ boson exchange has been excluded by direct detection experiments.
This connection can be avoided for other types of scalar dark matter. 
If there is a mass splitting between the scalar and pseudo-scalar component of $S$, the lighter one will be the DM candidate and its scattering off a nucleus via $Z$ boson exchange can be kinematically forbidden, provided that its kinetic energy is less than the mass difference between scalar and pseudo-scalar component of $S$. As the average velocity of dark matter during freeze-out is much larger ($(v/c)^2\sim 1/20$) compared to the average local velocity of dark matter ($v\sim 220\mathrm{km}/\mathrm{s}$), coannihilation via $s$-channel $Z$ exchange still occurs.
Another possibility is to introduce another $S^\prime$, a singlet under SU(2)$\times$U(1)  with the same
$G_\nu$ quantum numbers as $S$. We can
write a term of form $S' H^\dagger\cdot S$ which leads to a mixing
between $S$ and $S'$. The dark matter will be the lighter combination of $S$ and $S'$ and its annihilation cross section will then be given by
the same formula as $\sigma(S\bar{S}\to f \bar{f})$  rescaled by a factor of $\sin^4\alpha$
 where $\alpha$ is the mixing.  Thus, by adjusting $\alpha$,
 $\sigma (S\bar{S}\to f\bar{f})$  can be tuned and lead to the correct relic abundance.
For $m_b<m_{DM}<m_W$, a mixing angle $\sin\alpha=0.5$ and taking a typical velocity at freeze-out of $(v/c)^2 \sim 1/20$, we estimate for $m_S=60$~GeV,
\begin{equation}
 \langle \sigma_{tot} v\rangle\simeq 3\times 10^{-26}~ \frac{{\rm cm}^3}{\rm s} \frac{(m_S/60~{\rm GeV})^2}{\left(1-4 (m_S/60~{\rm GeV})^2/m_Z^2\right)^2} \left(\frac{\sin\alpha}{0.5}\right)^4\;. 
\end{equation}
 For higher values of $m_S$, $\langle \sigma_{tot} v \rangle$ can exceed
$3\times 10^{-26}~{\rm cm}^3/\mathrm{s}$ especially when new annihilation modes to $t
\bar{t}$, $W^+W^-$ and $Z Z$ open up. For $m_S\gg m_{EW}$, we can
write
\begin{equation}
\langle \sigma_{tot} v
\rangle\simeq 3\times 10^{-26}~\frac{{\rm cm}^3}{\rm s} \left(\frac{1.1~{\rm
TeV}}{m_S}\right)^2 \left(\frac{\sin\alpha}{0.5}\right)^4\;.
\end{equation}

If the dark matter is fermionic and belongs to an SU(2) doublets, the decay  channel through  $Z$ boson exchange is open with cross section
\begin{equation}
\langle \sigma(F\bar F\to f\bar f)v \rangle =\frac{4N_c G_F^2  }{
  \pi}\frac{(|a_L|^2+|a_R|^2) M_F^2}{(1-4m_F^2/m_Z^2)^2}.
\end{equation}
Unlike the previous case, there is  no  $v^2$ suppression factor, since the initial state particles are Dirac fermions rather than scalars and the annihilation can be $s$-wave.
 The cross-section  will then exceed $3\times 10^{-26}~{\rm cm}^3/{\rm s}$ for
  $1~{\rm GeV}<m_F<1~{\rm TeV}$. 
If the dark matter is fermionic and belongs to an SU(2) doublet, the annihilation cross section of dark matter via $Z$ bosons is directly related to the direct detection cross section. Even an annihilation cross section of $\ev{\sigma v}\sim 3\times 10^{-26} {\rm cm}^3/\mathrm{s}$ leads to a direct detection cross section of the order of $10^{-39} \mathrm{cm}^2$ for fermionic DM, which has been excluded by direct detection experiments (see e.g.~\cite{Aprile:2012nq}). Hence, the annihilation via $s$-channel $Z$ boson exchange can only lead to a subdominant contribution of the DM annihilation cross section. 
This bound can be avoided by introducing a singlet  $F'$  that mixes with $F$, such that the annihilation cross section is reduced by the mixing analogously to the scalar dark matter case.

\mathversion{bold}
\subsection*{Annihilation via $t$-channel $F$ ($S$) exchange}
\mathversion{normal}
Let us consider the case in which $S$ is a singlet that plays the role of dark matter.
The annihilation to $l \bar{l}, \nu \bar{\nu}$ via $t$-channel $F$ exchange is helicity suppressed and cannot account for the  required total annihilation rate. However, the related three-body decay with the additional emission of a gauge boson, like electromagnetic\cite{Bergstrom:1989jr} or electroweak\cite{Bell:2010ei,*Bell:2011if} bremsstrahlung can account for the thermal DM annihilation cross section. 
The annihilation to $\nu\nu$ via the helicity flipping $t$-channel $F$ exchange is suppressed by a lepton number violating coupling and suppressed by the mass of the exchanged fermion $F$, which both also control the smallness of neutrino masses. The cross section ends up to be too small for most regions of parameter space~\cite{Farzan:2010mr}, see however \cite{Ma:2006km,Boehm:2006mi}.

Let us finally discuss if the dark matter is fermionic and belongs to an SU(2) doublets, the dominant annihilation modes can be
$F\bar{F} \to \nu \bar{\nu}, l \bar{l}$ via $t$-channel $S$ exchange.
In this case, there is no $p$-wave suppression
and the cross section can be of order  of
\begin{equation}
\ev{ \sigma(F\bar{F} \to  \nu \bar{\nu}, l \bar{l})v}=
 \frac{g^4}{8\pi}
\frac{m_F^2}{(m_F^2-m_S^2)^2}\;,
\end{equation}
neglecting the final state masses. Here $g$ denotes a generic Yukawa coupling defined in \Eqref{L-Y}.  
Taking $m_S\sim m_F\sim 100$~GeV, $g$ should be of the order of 0.1.
With such a large coupling,  $\mu \to e \gamma$ exceeds the experimental bounds unless a specific flavor structure is invoked to suppress this process.

\mathversion{bold}
\subsection*{Annihilation via $s$-channel Higgs exchange}
\mathversion{normal}
Finally, we give the annihilation via Higgs exchange. Away from the resonant production of the Higgs, the annihilation of $S$ via $s$-channel can be related to the Higgs decay width for a Higgs into a final state $X$
\begin{align}\label{eq:HiggsMediated}
\braket{\sigma(SS\to h^*\to X)_H v}&= \left.(2m_h\Gamma(h\to X))\right|_{m_h\to2 M_S} \frac{1}{4M_S^2}\frac{4|\lambda v|^2}{(4M_S^2-m_h^2)^2}\\\nonumber
&= \frac{\left.\Gamma(h\to X)\right|_{m_h\to2 M_S}}{M_S}\frac{4|\lambda v|^2}{(4M_S^2-m_h^2)^2}\; ,
\end{align}
with the Higgs mass $m_h$ and the effective coupling  of $S$ to the Higgs $h$ defined by $\mathcal{L}\supset(\lambda v) h SS$. The DM phenomenology is similar to a scalar singlet DM model (see e.g.~\cite{Silveira:1985rk}). 
In particular, there will be a close correlation between annihilation rate of $S\bar S$ and the DM-nucleon cross section and therefore the direct detection rate. 

Similarly to \Eqref{eq:HiggsMediated}, we can calculate the $S$-wave contribution to the annihilation of fermionic dark matter $F$ via $s$-channel Higgs $h$ exchange into a final state $X$
\begin{align}
\braket{\sigma( F\bar F\to h^*\to X)_H v} &= \left.(2m_h\Gamma(h\to X))\right|_{m_h\to2 M_F} 
\frac{1}{4 M_F^2} \frac{2 M_F^2 \left(|y_L|^2+|y_R|^2\right)}{(4 M_F^2-m_h^2)^2}\;,
\end{align}
with the Yukawa coupling $\mathcal{L}\supset - \bar F (y_L P_L +y_R P_R) h F$ coupling to the CP even scalar $h$, which leads to a similar phenomenology like the fermionic singlet DM model (see e.g.~\cite{Kim:2008pp}). As for the scalar case, a relation between the annihilation rate and the direct detection is present.

\section{Conclusions and discussion\label{sec:con}}

The smallness of neutrino masses is one of the longstanding
problems in the phenomenology of particle physics. Various models
have been proposed in the literature within which neutrino
masses are produced at loop level so their smallness is natural. In this paper, we have discussed the class of models
within which neutrino masses are produced at loop level via a
Yukawa term that couples neutrinos to new scalars and
fermions. We studied and outlined some general results that can be
drawn from the topology of neutrino mass diagrams or the symmetry,
$G_\nu$, imposed on the model. We have discussed conditions on the
$G_\nu$ symmetry and topology of loop diagrams that forbid the
presence of a lower order, and consequently dominating,
contribution to the neutrino mass. Under these conditions these
diagrams will therefore give the dominant contribution to the
neutrino mass. More general results are outlined item by item in
sect.~\ref{sec:generalisation}.

In this paper, we have assumed that the $G_\nu$ symmetry remains
unbroken. In case that the SM particles are invariant under the
$G_\nu$ symmetry, the lightest new particle with a non-trivial
behavior under this symmetry will be stable. If this particle is
neutral, it can contribute to the dark matter in the universe.
Independently of a given model, there have been studies of the
impact of different symmetries on the DM predictions (See
e.g.~\cite{Ma:2007gq,Batell:2010bp,Adulpravitchai:2011ei,Belanger:2012vp}).
We briefly discuss the implications of the discrete
symmetry for dark matter stabilization and discuss the possibility
of the existence of multiple dark matter candidates for an abelian group
that can be decomposed to the direct product of other groups.
We also discussed various possible modes of annihilation of a dark
matter pair.

Within the class of models that we have discussed in this paper,
the scale of new physics can be as low as the electroweak scale.
The new particles that are added have no strong interactions;
however, they can have electroweak interactions. The model can
include new charged particles coupled to the Higgs which, along
with the SM contributions, may explain the possible excess in the
diphoton Higgs decay channel~\cite{Carena:2012xa}.

 At a hadron
collider such as the LHC, the only production mode of these
particles is through electroweak interactions but in a lepton
collider such as the ILC, these particles can be also produced via
Yukawa interactions in Eq.~(\ref{L-Y}) in the $t$-channel. If all
the SM particles transform trivially under $G_\nu$ and all new
particles carry $G_\nu$ charges,  these particles can be produced
at colliders only in association with other new particles such
that the final products form a singlet of $G_\nu$. For example, if
$G_\nu=Z_2$, the new particles that are odd under $Z_2$ can be
produced only in pairs.
Moreover, the $G_\nu$ symmetry implies that the decay products of
new particles include lighter new (beyond SM) particles. In fact
these particles will go through a chain of successive decays until
they produce stable new particle(s). If these final stable
products are neutral, they will appear as missing energy but, if
they are charged, they can be detected. Since they are heavy and
have no strong interactions, they will generally lose energy with a
rate smaller than the muon energy loss rate which means they come
to rest only after they exit the detector. In this
case, the signature of the model will be quite distinct from the
SM background, raising the discovery chance of the model.

Decay of the new particles can take place through the Yukawa
coupling in \Eqref{L-Y} which means each decay produces a
lepton along with a new particle. The branching ratio to different
flavors is determined by $g_{ij\alpha}$. This is the same coupling
that determines the flavor structure of the neutrino mass matrix.
In principle, by studying the flavor composition of the decay
products of the new particles, one can cross check these models.
This possibility has been studied in detail in
\cite{Farzan:2010fw} for the specific case of the SLIM model
\cite{Boehm:2006mi,Farzan:2009ji} where $G_\nu=Z_2$ and the new
fermions are neutral. Unfortunately, the high rate of background
and the uncertainty in luminosity will limit the capability of the
LHC to extract the flavor structure of the coupling. However, a
lepton collider can have a better chance of determining the
coupling. If the model contains multiply charged particles, their
production will be enhanced by square of their charge. Moreover
their successive decay to multiple charged leptons plus missing
energy or new stable charged particle will provide a distinct
signature enhancing the chances of discovery at lepton and hadron
colliders.

One of the key ingredients used in our setup to suppress lower loop contributions is the presence of discrete symmetries at quantum level. Although in our
setup discrete symmetries are global and not gauged, there are arguments that all symmetries, including discrete global symmetries, should be gauged
in a theory of quantum gravity \cite{Ibanez:1991hv,*Banks:1991xj}. This is due to the expectation, mainly advocated in string theory, that all symmetries have a geometric origin and the space-time itself is a locally constructed, secondary concept, in these settings. Thus, all the symmetries should be local symmetries. These symmetries could be broken due to quantum gravity effects and/or anomalies~\cite{Araki:2006mw,*Araki:2008ek}.  The effects of symmetry
breaking (if there is any) are expected  to be suppressed by
inverse powers of $m_{Pl}$. Based on the dimensional analysis, we
can estimate the contribution from the $G_\nu$ violating effects
to the neutrino mass to be at most $ (\ev{H}^2/m_{Pl}) (M_{NEW}/m_{Pl})^n\sim
10^{-6} \eV ( M_{NEW} / m_{Pl} )^n$ for some $n=0,1,2,\dots$. Thus, we can safely
neglect this effect.

In summary, small neutrino masses can be generated at the loop level
in models in which the leptons couple to the new sector.
New symmetries guarantee the absence of a Dirac mass term for neutrinos
and can forbid lower loop diagrams. The presence of this symmetry
leads to neutral stable candidates which might explain the observed
dark matter abundance in the Universe.
The additional suppression due to the higher loop order
allows to lower the scale of new physics down to the TeV scale
keeping large couplings and providing specific testable signatures at colliders
and observable lepton flavor violating processes.

%%%%%%%%%%%%%%%%%%%%%%%%%%%%%%%%%%%%%%%%%%%%%%%%%%%%%%%%%%%%%%%%%%%%%%%%%
%%%%%%%%%%%%%%%%%%%%%%%%%%%%%%%%%%%%%%%%%%%%%%%%%%%%%%%%%%%%%%%%%%%%%%%%%
%%%%%%%%%%%%%%%%%%%%%%%%%%%%%%%%%%%%%%%%%%%%%%%%%%%%%%%%%%%%%%%%%%%%%%%%%

\section*{Acknowledgements}
The authors acknowledge partial support from the  European Union FP7  ITN INVISIBLES (Marie Curie Actions, PITN- GA-2011- 289442). They also thank Galileo Galilei Institute for Theoretical Physics for its hospitality.
YF is grateful to ICTP for partial financial support and hospitality. She thanks M. M. Sheikh-Jabbari for useful discussions.
MS acknowledges support by the Australian Research Council.

%%%%%%%%%%%%%%%%%%%%%%%%%%%%%%%%%%%%%%%%%%%%%%%%%%%%%%%%%%%%%%%%%%%%%%%%%
%%%%%%%%%%%%%%%%%%%%%%%%%%%%%%%%%%%%%%%%%%%%%%%%%%%%%%%%%%%%%%%%%%%%%%%%%
%%%%%%%%%%%%%%%%%%%%%%%%%%%%%%%%%%%%%%%%%%%%%%%%%%%%%%%%%%%%%%%%%%%%%%%%%

\bibliography{draft}

\end{document}